\documentstyle[aps,epsfig,psfig]{revtex}      
 
\begin{document} 
\title{Scaling limit of virtual states of triatomic systems} 
\author{ M. T.  Yamashita$^{1}$,  T. Frederico$^{2}$, A.Delfino$^{3}$, 
Lauro Tomio$^{4}$} 
\address{
$^{1}$ Laborat\'orio do Acelerador Linear, 
 Instituto de F\'{\i}sica, Universidade de S\~{a}o Paulo,
\\
C.P. 663118, CEP 05315-970, S\~{a}o Paulo, Brasil \\
$^{2}$ Dep. de F\'\i sica, Instituto Tecnol\'ogico de Aeron\'autica,
Centro T\'ecnico Aeroespacial,\\ 
12228-900 S\~ao Jos\'e dos Campos, Brasil \\
$^3$Departamento de F\'\i sica, Universidade Federal Fluminense,
\\ 24210-340  Niter\'oi, Rio de Janeiro, Brasil.\\
$^{4}$ Instituto de F\'\i sica Te\'orica, Universidade Estadual
Paulista, 01405-900 S\~{a}o Paulo, Brasil \\ } 
\date{\today} 
\maketitle 
\begin{abstract} 
For a system with three identical atoms, the dependence of the $s-$wave virtual
state energy on the weakly bound dimer and trimer binding energies is calculated in
a form of a universal scaling function. The scaling function is obtained from a
renormalizable three-body model with a pairwise Dirac-delta interaction. 
It was also discussed the threshold condition for the appearance of the trimer 
virtual state.
\end{abstract} 
\vskip 0.5cm 

\section{Introduction}
 
Weakly bound three-body  zero-angular momentum states appear in a three boson
system, with the number of states  growing to infinity, condensing at zero energy as
the pair interactions are just  about to bind two particles in $s-$wave. These
three-body states are known as Efimov states~\cite{ef70,ef90}. Their wave functions,
loosely bound, extend far beyond those of normal states and dominate the low-energy
scattering phenomena in these systems. The Efimov states have been studied in a
number of model calculations~\cite{fe93,efim2,tom}, 
in atomic and nuclear systems,
without yet a clear  experimental signature of their 
occurrence~\cite{ef90,evid1,am92,am97,am99,de00}. 

Actually, the search of Efimov states in atomic systems is becoming more appealing,
due to the experimental realization of Bose-Einstein condensation
(BEC)~\cite{bec95}, and due to the possibility of altering the effective scattering
length of the low-energy atom-atom interaction in the trap, from large negative to
large positive values crossing the dimer zero binding energy value, by using an
external magnetic field~\cite{avaria}.  This possibility of changing the two-body
scattering length to large values, as recently shown in Ref.~\cite{JILA02},
can alter in an essential way the balance between the non-linear
first few terms of the mean-field description presented in the equations that model
Bose-Einstein condensed gases~\cite{gammal}. This can certainly open new perspectives
for theoretical and experimental investigations related to the many-body behavior of
condensate systems. Even in systems where the occurrence of an excited bound Efimov
state has shown to be doubtful or even not possible, as for example, in the case of
halo nuclei like $^{20}$C or $^{18}$C (seen as a core with a halo of two
neutrons)~\cite{am97}, one can verify the occurrence of three-body virtual states. 
The physics of these three-body systems is related to the unusually large size of
the wave-function compared to the range of the potential. Thus, the detailed form of
the short-ranged potential is not important for the three-body  
observables\cite{jensen}, which gives to the system universal properties, defined by
few physical scales\cite{am97}. Strictly speaking, in the limit of a zero-range
interaction the three-body system is parametrized by the physical two- and
three-body scales, which are identified with the two-body scattering lengths and  one
three-body binding energy~\cite{am99,ad95}. The physical reason for the sensibility
of the three-body binding energy to the interaction properties comes from the
collapse of the system in the  limit of a zero-range force, which is known as the 
Thomas effect~\cite{th35}. 

In the present work, we analyze the possibility that an excited trimer state
becomes a virtual state, when the physical scales of the system are changed.
This is expected to occur, for example, near the limit when the two-body scattering
length goes from large positive to large negative value: the corresponding two-body
energy is close to zero and goes from a bound to a virtual state, with appearance
of many  bound and virtual three-body states.
The three-body virtual state energy is a pole of the S-matrix in the 
second sheet of the complex energy plane. In a general case, as the strength 
of the two-body potential diminishes, the pole moves towards the first 
energy sheet to become a bound state~\cite{tom}. 
More recently, this behavior of the Efimov state going to a
virtual state with the increase of the strength of the interaction,
has been confirmed in realistic calculation of the helium trimer~\cite{kol99}. 
Here, we study a new physical aspect of the emergence of the $s-$wave 
virtual state from an Efimov state: {\it it appears when   
the ratio between the dimer and trimer binding energies grows.}
This approach goes beyond a previous analysis of excited three-body bound states
with short-range interactions, that was performed in Ref.~\cite{am99}. 
In Ref.~\cite{am99}, a scaling function was introduced to analyze the behavior
of bound Efimov states when modifying the triatomic physical scales. 
Essentially, we are extending to the second sheet of the complex energy
plane (to include virtual trimer states) a previous investigation on a
universal scaling mechanism that was applied to two and three-body 
bound states\cite{am99,de00}. 
The extension of the scaling function to the second energy sheet is performed by
following the Efimov states as they move from bound to virtual, accordingly to the
variation of the ratio of the dimer to trimer bound state energies.
On the other hand, as we present the discussion 
through a universal scaling mechanism with the results in 
dimensionless units, all the conclusions apply equally to any 
low-energy three-boson system. For the regularization and 
renormalization of the zero-range model, we 
compare two different approaches: by using a momentum 
cutoff parameter~\cite{am99} and via kernel 
subtraction~\cite{ad95,fre00,fix}. As the two-body energy 
goes to zero (or equivalently the regularization parameter goes
to infinity), we conclude that the results of both methods do not differ. 

The paper is organized as follows. 
In section II, we generalize the scaling function defined in Ref.~\cite{am99} 
to include virtual trimer states. In this section, we also
revise the connection between the Thomas and Efimov effects, while 
introducing our notation and the homogeneous integral equation for the Faddeev
component of the vertex of the wave-function for zero-range potential. 
In section III we present our main numerical results. In the first
subsection, we present the subtracted homogeneous Faddeev 
equation that we have
used for determining the trimer bound and virtual states and we
briefly explain how the renormalization method of
Refs.~\cite{fre00,fix} implies in the subtracted three-body equation 
first formulated in Ref.~\cite{ad95}.
In the next subsection, we
present our new numerical results for the virtual state energies,
including the bound-state previous results and we compare, as well, 
the results obtained by using the sharp-cutoff and the subtraction scheme. 
Comparison with other calculations are also discussed.
Our conclusions are summarized in section IV.

\section{Thomas-Efimov Effect and the Generalized Scaling Function }

In this section, we introduce the generalization of the scaling 
function defined in Ref.~\cite{am99}, to be used in the second energy 
sheet of the trimer energy. In order to become clear this extension, 
and to define our notation, we begin by revising the main findings
of Refs.~\cite{am99,adh88}.

The two-boson system in the limit of a zero-range interaction has only 
one physical scale, that one can choose the scattering length ($a$) or the 
energy of the bound or virtual state.  The two-body $s-$wave scattering 
amplitude in units  of $\hbar=m=1$ is parametrized as a function of the 
momentum $k$, by $f(k)=\left( k\cot\delta_0 -{\rm i}k\right)^{-1},$
where the $s-$wave phase shift $\delta_0$ is given by
$k\cot\delta_0 =-a^{-1}+\frac12 r_0 k^2 + ... ,$ and $r_0$ is the 
effective range. For $ a \ > \  0 $ the two-body system is bound, 
otherwise, for $a \  < \ 0$, it is virtual. A short range potential is 
characterized by $r_0|a|^{-1} << 1 $ and, in this case, 
$f(k)=\left( -a^{-1}- {\rm i}k \right)^{-1}$ and $a^{-1}=\pm\sqrt{E_{2}}$ 
($+$ for bound and $-$ for virtual state). 
 
The three-boson system for $\ell=0$ in three dimensions collapses
when $r_0\rightarrow 0$ with a fixed two-body scale, which is  
known as Thomas effect\cite{th35}. Thus, the 
three-body system has a characteristic physical scale 
independent of the two-body ones\cite{ad95}. In one and two
space dimensions the collapse is absent\cite{jensen}. 
In the limit when the binding energy of the two-boson system goes
to zero, the three-boson system has an infinite number of bound Efimov 
states~\cite{ef70} condensing at zero-energy. 
The Thomas and Efimov effects were shown to be physically 
equivalent~\cite{adh88}, since in both cases the ratio between the 
interaction range and the two-body scattering length goes to zero.

The integral equation for the Faddeev components, $\phi$,
of the three boson bound state vertex, for $\ell=0$, with the zero-range 
interaction,  needs a momentum cut-off $(\Lambda$) of the order of 
$r_0^{-1}$, due to the Thomas collapse. 
According to Ref.\cite{adh88}, using units of $\Lambda=1$, we 
rescale the momentum variables and the two and three-body binding energies, 
respectively, such that $ \vec p = \Lambda \vec x$, $\vec q = \Lambda 
\vec y$, $E_2 = \Lambda^2 \epsilon_{2}$, and
$E_3= \Lambda^2 \epsilon_{3}$. In this dimensionless variables, after
redefining $\phi$ as $\chi(\vec x)\equiv \Lambda^{3/2}\phi(\vec p)$,
we obtain the integral equation~\cite{adh88,am92,am97}:
\begin{eqnarray}
\chi(\vec y)=\frac{-\pi^{-2}}{ \pm \sqrt{\epsilon_2} 
-\sqrt{\epsilon_3+\frac34 \vec y^2} }
\int d^3x \; 
\frac{\theta(1-|\vec x|)}{\epsilon_{3}+\vec y^2+\vec 
x^2+\vec y .{\vec x}} \;\chi(\vec x) \ .\label{skt}
\end{eqnarray}
The number of three-body bound states, given by the values of $\epsilon_3$ 
that satisfies Eq. (\ref{skt}), grows without limit when $\epsilon_2$ 
decreases to zero: $\epsilon_3 = \epsilon_3^{(N)}$ $(N=0,1,2...)$, 
with $\epsilon_3^{(N)}/\epsilon_3^{(N+1)}\approx$ 500~\cite{ef70}. 
They are the energies of the Efimov states, in  units of $\Lambda=1$.
But, the limit of $\epsilon_2$ going to zero, can be realized either by 
$E_2\rightarrow 0$ (with a fixed $\Lambda$) or by 
$\Lambda\sim r_0^{-1}\rightarrow\infty$ (with $E_2$ fixed). 
In this last case, the range of the interaction is set to zero and  the 
system collapses: $E_3^{(0)} = \epsilon_3^{(0)}\Lambda^2 \to \infty$.
This is known as the Thomas collapse of the three-body ground state.
Therefore, the Thomas and the Efimov states are given by the same
limit $\epsilon_2\to 0$ of Eq.~(\ref{skt}), and are related 
by a scale transformation\cite{adh88}. 

Now, the concept of the scaling function is introduced according to
Ref.\cite{am99}. For a nonvanishing $\epsilon_2$, 
the solutions of Eq.~(\ref{skt}) defines the dimensionless three-body 
energies as functions of $\pm\sqrt{\epsilon_2}$:
$\epsilon_3^{(N)}\equiv
\epsilon_3^{(N)}\left(\pm\sqrt{\epsilon_2}\right) \ $.
Using the $N-$th energy to obtain $\Lambda$, then
$ \Lambda^2= E_3^{(N)}/\epsilon^{(N)}_3$,  and
\begin{eqnarray}
 E_3^{(N+1)}=E_3^{(N)}\ . \ 
\frac{\epsilon_3^{(N+1)}\left(\xi\right)}{\epsilon_3^{(N)}}\ ,
\label{e3}
\end{eqnarray}
where $\xi\equiv\pm\sqrt{\epsilon_2}=\pm
{\left(E_2\; \epsilon^{(N)}_3 / E^{(N)}_3 \right)^{1/2}}$.
In Eq.(\ref{e3}), the two and three body physical scales determines
$E^{(N+1)}_3$, the next excited state above $E^{(N)}_3$. 
In Ref.~\cite{am99}, $E^{(N)}_3$ was identified with the three-body 
scale, as any state $N$ works equally well to set the trimer scale.
However, we will be interested in the two most excited
three-body states, that in practice we are going to identify with the
ground and first excited state in triatomic systems.
This identification is unambiguous because,
with $N$ and $N+1$  two consecutive excited states, 
the limit
\begin{eqnarray}
 \frac{E_3^{(N+1)}}{E_3^{(N)}}\ = \ \lim_{N\rightarrow\infty} 
\frac{ \epsilon_3^{(N+1)}\left(
\xi\right)}
{\epsilon_3^{(N)}}
\ = \ {\cal F}\left(\pm\sqrt{\frac{E_2}{E^{(N)}_3}}\right)
\label{lim}
\end{eqnarray}
exists and defines the scaling function ${\cal F}$\cite{am97,am99}. 
A qualitative argument to explain the scaling limit has been
provided in Ref.~\cite{am99} based on the notion of the 
long-range potential \cite{ef70,ef90,fonseca}.

In the next, we provide the 
generalization of the scaling function (\ref{lim}), that is obtained
by extending the formalism to the second sheet of the three-body complex 
energy plane. In the present approach, we only consider the two-body 
subsystem as bound. For this purpose, we define the general scaling 
function ${\cal K}$, given by
\begin{eqnarray}
\ {\cal K}\left(\sqrt{\frac{E_2}{E^{(N)}_3}}\right)\ 
\equiv \pm \sqrt{\frac{E_3^{(N+1)}-E_2}{E_3^{(N)}}}\
= \pm \sqrt{\frac{\epsilon_3^{(N+1)}-\epsilon_2}{\epsilon_3^{(N)}}}\
\label{lim1}.
\end{eqnarray}
This defined function ${\cal K}$ has its values on the imaginary
axis of a three-body momentum space; a space that is defined with origin 
at the point in which the energies of the three-body system and the bound 
two-body subsystem are equal($E_3=E_2$). In this respect, relative to the 
bound subsystem, we can define bound and virtual states for the three-body 
system: ${\cal K}$ assumes a negative value for a three-body virtual 
state; and a positive value for a three-body bound-state.  
Schematically, we represent in Fig.~1 the energies of the two- and three- 
body system in the complex energy plane. The two-body subsystem is bound
and the three-body system can be bound or virtual, with the energies 
given, respectively by $\epsilon_{3B}$ and $\epsilon_{3V}$.
Through the elastic cut (corresponding to the atom-dimer elastic 
scattering) one defines two-sheets; in the first sheet, 
we have the three-body bound 
state energy at Re$(\epsilon)=-\epsilon_{3B}$;
in the second sheet, the three-body virtual state energy at 
Re$(\epsilon)=-\epsilon_{3V}$, as illustrated in Fig.~1.

\begin{figure}[thbp]
\centerline{\epsfig{figure=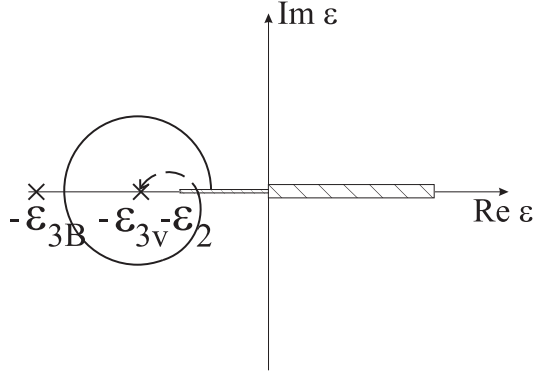,width=7cm}}
\caption{Schematical representation of the complex energy plane, in our 
dimensionless units. $\epsilon_{3B}$ and 
$\epsilon_{3V}$ are, respectively, pictorial representations of the
positions of the three-body bound- and virtual-state energies in the 
first and second three-body energy sheet. 
The three-body cut is shown for Re$(\epsilon) > 0$. The 
elastic cut (the narrow one) is shown with the origin at 
Re$(\epsilon)=-\epsilon_2$, where $\epsilon_2$ is the energy of the
two-body bound state. 
} 
\label{fig1}
\end{figure}

We would like to add one more comment to this section.
The existence of a three-body scale implies in
the low energy universality found in three-body systems, 
or correlations between three-body observables~\cite{fre87a,ad95}.
In the scaling limit, one has

\begin{eqnarray}
{\cal{O}}\left(E, E_{3},E_{2}\right)=(E_{3})^\eta
{\cal A}\left( \sqrt{E/E_{3}},\sqrt{E_{2}/E_{3}}\right) \ ,
\label{o}
\end{eqnarray}
where $\cal O$ is a general observable of the three-body system at energy $E$,
with dimension of energy to the power $\eta$.
The scattering amplitude of the elastic process a + bc $\rightarrow$ a + bc,  
$ f_{3}= \sqrt{E_{3}}^{-1}\;F( \sqrt{E/E_{3}},
\sqrt{E_{2}/E_{3}})$
for $E=E_2$,  implies that the scattering length is given by a function
$a_{3}= \sqrt{E_3}^{-1} \; F( \sqrt{E_2/E_3})$. 
In the  three-nucleon system  this  originates
the ``Phillips plot'', the correlation between the doublet neutron-deuteron 
 scattering length and the triton  energy~\cite{phillips}. 
The scaling functions Eqs. (\ref{lim}) and (\ref{lim1}) express the
correlation between the excited or virtual state energies of the trimer
and its ground state energy, which can be understood as  particular cases of 
Eq. (\ref{o}). 

\section{Numerical Results for Virtual and Bound Trimers}

In this section we present our main results for the trimer 
bound and virtual states.
With the sake to be complete, we first briefly sketch a new derivation 
of the subtracted equations that were numerically solved.

\subsection{Renormalization and Subtracted Equations}

The homogeneous form of the subtracted 
Faddeev equation~\cite{ad95} for the bound three-boson 
system with a zero-range interaction, is given by:
\begin{eqnarray}
\chi(\vec y)=\frac{-\pi^{-2}}{ \pm \sqrt{\epsilon_2} 
-\sqrt{\epsilon_3+\frac34 \vec y^2} }
\int d^3x \left(
\frac{1}{\epsilon_{3}+\vec y^2+\vec x^2+\vec y .{\vec x}}
-\frac{1}{
1+\vec y^2+\vec x^2+\vec y .{\vec x}}\right) {\chi(\vec x)}\ ,
\label{skren0}
\end{eqnarray}
which is written in units such that the three-body 
subtraction energy is $\mu_{(3)}^2=1$.
It has a similar form as Eq.(\ref{skt}) with a different regulator,
that expresses the physical condition at the subtraction point.

We briefly explain below the main physical steps to derive the
three-body renormalized equation~\cite{ad95} used in our 
numerical calculation of the scaling functions through Eq.(\ref{skren0})
for the bound state  and its analytic continuation to the
second energy sheet for the virtual state. 
We begin from the general 
Lippman-Schwinger equation expressed in a subtracted form~\cite{fre00}: 
\begin{eqnarray}
T_{\cal R}(E)=T_{\cal R}(-\mu^2)+T_{\cal R}(-\mu^2)\
\left[G^{(+)}_0(E)-G_0(-\mu^2)\right]T_{\cal R}(E) , 
\label{tr}
\end{eqnarray} 
where $T_{\cal R}(-\mu^2)$, which is the T-matrix at a 
given energy scale $-\mu^2$ (negative energy, for 
convenience),
$G_0^{(+)}(E)=[E-H_0+{\rm i}\delta]^{-1}$ and $H_0$ is the free
Hamiltonian.
Equation (\ref{tr}) defines the renormalized T-matrix in which
$T_{\cal R}(-\mu^2)$ is known and replace the original
ill-defined potential $V$:
\begin{eqnarray}
T_{\cal R}(-\mu^2)=\left[1-VG_0(-\mu^2)\right]^{-1}V \ .
\label{v}
\end{eqnarray}
The renormalized T-matrix does not depend on the arbitrary 
subtraction point $-\mu^2$ (once ${dV}/{d\mu^2}=0$), 
which implies  in a Callan-Symanzik~\cite{fre00,fix}
type equation for $T_{\cal R}(-\mu^2)$:   
\begin{eqnarray}
\frac{d}{d\mu^2}T_{\cal R}(-\mu^2)=T_{\cal R}(-\mu^2)\left[G_0(-\mu^2)
\right]^2
T_{\cal R}(-\mu^2) \ .
\label{cs}
\end{eqnarray}
This expresses the renormalization group invariance of the 
subtracted equation. 

To solve Eq.(\ref{tr}) for the three-body T-matrix,
$T^{(3)}_{\cal R}(E)$, a dynamical assumption has to be made at 
a particular subtraction point $-\mu^2_{(3)}$, where 
we assume that the three-body T-matrix is equal to the driving term,
which is given by the sum of the pairwise two-body 
T-matrices. Thus, at the energy $-\mu^2_{(3)}$ it is assumed 
that the three-body multiple scattering series vanishes beyond
the driving term. Observe that this is not true
for a regular finite range potential, only  in the 
limit of $\mu_{(3)}\rightarrow \infty $. However, in the scaling limit, 
in fact the actual value of $\mu_{(3)}$ tends to infinity such that  
$E_2/\mu^2_{(3)}$ goes to zero, as it be will be clear in 
our numerical calculations. 

With our assumption, the T-matrix at the subtraction point $\mu_{(3)}$
is given by
\begin{eqnarray}
T^{(3)}_{\cal R}(-\mu_{(3)}^2)=\sum_{(ij)}
T^{(2)}_{{\cal R}(ij)}\left(-\mu_{(3)}^2-\frac{q^2_{k}}{2m_{k(ij)}}\right)
,\label{vrentres1}
\end{eqnarray}
where $(i,j,k) = (1,2,3),\; (2,3,1), \;(3,1,2)$.
The summation is performed over all pairs and the renormalized two-body 
T-matrix elements for the pair $(ij)$ are given by 
$\langle {\vec P}^\prime|T^{(2)}_{\cal R}(E)|\vec P\rangle  = 
1/\left[2\pi^2(\pm\sqrt{E_2}+i\sqrt{E})\right]\ $.
The argument of the two-body T-matrix is the center of mass pair energy, 
where $q_{k}$ is the Jacobi relative momentum canonically conjugated to 
the relative coordinate of the particle $k$ to the center of mass of the 
pair $(ij)$, and $m_{k(ij)}$ is the reduced mass.

Using Eqs. (\ref{tr}) and (\ref{vrentres1}) and after some straightforward 
manipulations, the equations for the Faddeev components of the T-matrix 
at the bound state pole give Eq.(\ref{skren0}), which has a natural 
 momentum scale  given by $\mu_{(3)}$. In 
principle $\mu_{(3)}$  can be varied without
changing the content of the theory as long as the  three-body 
T-matrix at the new subtraction-energy $-{\mu^2_{(3)}}$ is found from 
the solution of Eq.~(\ref{cs}) with the boundary condition 
Eq.(\ref{vrentres1}),
and {consequently} 
Eq.(\ref{skren0}) {should be} conveniently rewritten. In the scaling limit,
Eq.(\ref{skt}) and Eq.(\ref{skren0}) produce the same results
(as we are going to illustrate numerically), 
since they are solved for $\epsilon_2$ going to zero, and the detailed
form of the regularization implied in both equations 
is not important anymore. However, Eq.(\ref{skren0}), has  conceptual
and practical advantages over Eq.(\ref{skt}), namely it is explicitly 
renormalization group invariant and it is as well regularized.

To simplify the notation of Eq. (\ref{skren0}), we introduce 
another definition related to the two-body energy:   
$\kappa_2 \equiv \pm \sqrt{\epsilon_2},$ 
where $+$ refers to bound and $-$ to virtual two-body state-energies. 
After partial wave projection of Eq.(\ref{skren0}), 
the s-wave  integral equation for the three-boson system is: 
\begin{eqnarray} 
\chi_s(y)&=&\tau(y;\epsilon_3;\kappa_2) 
\int_0^\infty dx\;x^2 G(y,x;\epsilon_3)\chi_s(x) \ ,  
\label{chi1} 
\end{eqnarray} 
where 
\begin{eqnarray} 
\tau(y;\epsilon_3;\kappa_2)&=& 
-\frac{2}{\pi} 
\left[\sqrt{\epsilon_3+\frac{3}{4} y^2}-\kappa_2\right]^{-1}, 
\label{tau1} \\    
G(y,x;\epsilon_3)&=& \left(\epsilon_3-1\right)
\int^1_{-1}dz \frac{1
}
{\left(\epsilon_3+y^2+x^2+yxz\right)
\left(1+y^2+x^2+yxz\right)}
\label{G1}  
\end{eqnarray} 
For the $\ell$-th angular momentum three-body state, the Thomas collapse is forbidden
if $\ell > 0$; consequently, no regularization is required and the integration over 
momentum can be extended to infinity even in the limit $\mu_{(3)}\rightarrow 
\infty$. 
For $\ell >0$, the original Skornyakov and Ter-Martirosian 
equation~\cite{sk} 
is well defined and the three-body observables are completely determined 
by the two-body physical scale corresponding to $E_2$.
One finds examples of the disappearance of the dependence 
on the three-body scale in $p-wave$ virtual states, for the trineutron system 
when  $n-n$ is artificially bound~\cite{Glockle,delf96} and in 
three-body halo nuclei (represented as a core with a halo of 
two neutrons)~\cite{delhalo}.

The analytic continuation to the second energy sheet, 
of the scattering equations for separable potentials, is
discussed in detail by Gl\"ockle, in Ref.~\cite{Glockle}. 
In the particular case of the zero-range three-body model~\cite{sk}, 
it is also given in Ref.~\cite{virtual}.
On the second energy sheet, the integral equations are obtained by 
the analytical continuation  through the two-body elastic scattering cut 
corresponding to the atom-dimer scattering.
The elastic scattering cut comes through the pole of the atom-atom 
elastic scattering amplitude in Eq.(\ref{tau1}). 
We perform the analytic continuation of Eq. (\ref{chi1}) to the 
second energy sheet. 
By substituting the spectator function $\chi_s(y)$ by 
$\overline\chi_s(y)\equiv 
(\epsilon_{3v}-\epsilon_2+\frac{3}{4}y^2)
\chi_s(y) $, where
$\epsilon_{3v}$ is the modulus of the virtual state energy, 
the resulting equation in the second energy sheet is given by:
\begin{eqnarray} 
\overline\chi_s(y)&=&\overline{\tau} (y;\epsilon_{3v};\kappa_2)
 \frac{4\pi \kappa_{3v}}{3}
G(y,-i\kappa_{3v};\epsilon_{3v})\overline\chi_s(-i\kappa_{3v})
\nonumber \\
&+&\overline{\tau}(y;\epsilon_{3v};\kappa_2)
\int_0^\infty dx\; x^2
 \frac{ G(y,x;\epsilon_{3v}) \overline\chi_s(x)}{\epsilon_{3v}-
\epsilon_2+\frac{3}{4}x^2} , 
\label{vchi1}  
\end{eqnarray} 
where the on-energy-shell momentum at the virtual state is
$\kappa_{3v}\equiv\sqrt{\frac{4}{3}(\epsilon_{3v}-\epsilon_2)}$ and
\begin{eqnarray} 
\overline{\tau}(y;\epsilon_{3v};\kappa_2)&\equiv&
-\frac{2}{\pi}
\left[\sqrt{\epsilon_{3v}+\frac{3}{4} y^2}+\kappa_2\right] \ .
\label{vtau2} 
\end{eqnarray} 

The cut of the elastic amplitude given by the exchange of one atom
between the different possibilities of the bound dimer
subsystems is near the physical region 
due to the small value of $\epsilon_2$. This cut is given by the values 
of imaginary $x$ between the extreme poles of the free 
three-body Green's function, $G(y,x;\epsilon_{3v})$, given by
Eq.(\ref{G1}) which appears in the right-hand-side of
Eq.(\ref{vchi1}),
\begin{eqnarray}
\epsilon_{3cut}+y^2+x^2+xyz = 0,
\label{cut}
\end{eqnarray} 
with $-1<z<1$, $y=x=-i \kappa_{cut}$ and
 $\epsilon_{3cut}=\frac{3}{4} \kappa_{cut}^2+\epsilon_2$. 
With the above, the cut satisfies 
\begin{eqnarray}
4\epsilon_2 \ > \ \epsilon_{3cut} \ > \ \frac43\epsilon_2 \ .
\label{cut1}
\end{eqnarray} 
The virtual state energy $\epsilon_{3v}$ in the second energy sheet is
found  between the scattering threshold and the cut,
$\epsilon_2 < \epsilon_{3v} <  \frac43 \epsilon_2$. 

\subsection{Scaling Plots}

It is usual to analyze how the Efimov states arise by varying the
strength of the interaction to change the value of the 
two-body binding energy. In our case, instead of this procedure,
we change directly the value of the energies in units of $\mu=1$ and 
by doing this we calculate the $s-$wave three-body energy evolution 
in the complex energy plane, corresponding to the bound and 
virtual triatomic states from Eqs. (\ref{chi1}) and 
(\ref{vchi1}), respectively. As $\epsilon_2$
goes to zero a crescent number of weakly bound (in units of $\mu=1$) 
Efimov states appear. 
The Thomas-Efimov limit for $\epsilon_2$ going to zero is clearly 
seen in Fig. 2, where we plot $\epsilon_3^{(N)}$ as a function of $\epsilon_2$.
In this figure we display only the energies of the first three states.
The main purpose of Fig. 2 is to show the real nature of the energies of 
the Thomas-Efimov states. The small circles and triangles correspond,
respectively, to the first and second excited virtual state energies, 
which begin at the cut from the one-particle exchange mechanism 
that gives $\epsilon_{3v}=(4/3) \epsilon_2$ (shown in the figure
by the dotted line). The threshold, from which the virtual three-body 
states arise, are exhibited by down-arrows $(\downarrow)$. 
When the two-body energy is enough for a trimer bound state to exist, 
then a decrease in $\epsilon_2$ allows the virtual state to appear from the 
one-particle-exchange cut. 
Further decrease in $\epsilon_2$ favors the appearance of the
excited state, which  emerges from the second energy sheet to the 
first one at the threshold value $\epsilon_3=\epsilon_2$, 
(solid line), indicated by the up-arrow $(\uparrow)$.
The critical value of $\epsilon_2$  is given by the ratio
$(\epsilon_2/\epsilon_3^{(N)})^{1/2}=0.38$ where the excited state
is labeled by $N+1$, and in the figure is indicated by the up-arrow.
This figure also strongly suggests that the Thomas-Efimov states cannot be 
completely understood only through the absolute value of $E_2$ itself, 
because the critical value for the appearance of the $(N+1)$-excited 
state depends only on the ratio 
$E_2/E_3^{(N)}=\epsilon_2/\epsilon_3^{(N)}$, 
which is independent of the absolute scale. 
Therefore, to show that this argument is universal, we study the function 
$E_3^{(N+1)}/E_2=\epsilon_3^{(N+1)}/\epsilon_2$ as a function
of $E_2/E_3^{(N)}=\epsilon_2/\epsilon_3^{(N)}$, where the $(N+1)$ state 
can be virtual or bound. This study is presented in Fig. 3.

\begin{figure}[h]
\centerline{\epsfig{figure=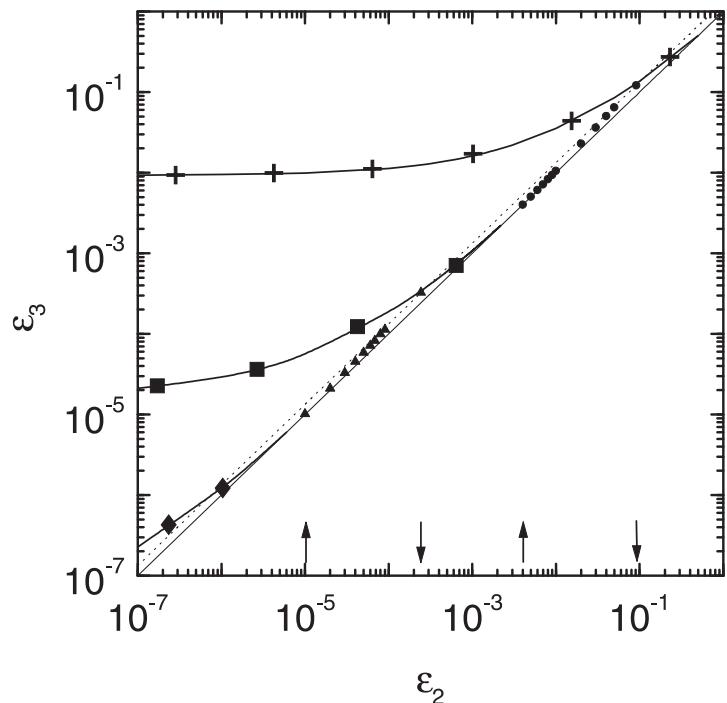,width=9.5cm}}
\caption{Trimer energies, $\epsilon_3$, as functions of the dimer bound
state energy $\epsilon_2$. The trimer ground state energy 
($\epsilon_3^{(0)}$) is shown by the curve with crosses; 
the first excited bound state ($\epsilon_3^{(1)}$) is shown by 
the curve with squares; and the second excited bound state ($\epsilon_3^{(2)}$) 
by the curve with diamonds. The behavior of two trimer virtual state 
energies, $\epsilon_{3v}^{(1)}$ (small circles) and $\epsilon_{3v}^{(2)}$ 
(small triangles) are also shown, as functions of the two-body energy, 
varying from  the threshold $\epsilon_3=\epsilon_2$ (solid line) to 
the threshold for the one-particle-exchange cut
$\epsilon_{3}=\frac43\epsilon_2$ (dotted line).
All the energies are given in arbitrary units.}
\label{fig2}
\end{figure}

The plot of Fig. 3 is constructed with the results for the first 
and second Thomas-Efimov states. 
This plot practically coincides with the corresponding one obtained
from the second and third states (not shown).
Figure 3 shows a universal route for the energy of the $(N+1)$ trimer state 
in the complex energy plane, from the second energy sheet to the first one 
as the ratio $E_2/E_3^{(N)}=\epsilon_2/\epsilon_3^{(N)}$ decreases. 
The three-body virtual state energy reaches $4E_2/3$ at 
$E_2/E_3^{(N)}=0.71$. Also realistic calculations
for the helium-trimer are available and are displayed in this figure. 
The agreement between our  calculations and the realistic ones, showns
the power of our scaling picture. Unfortunately, there is not yet, to
our knowledge realistic calculations of the virtual state in helium trimer or
even in any other weakly bound three-boson system, in which our route
should also applies. We emphasize that although we have presented results
only for the second and third Thomas-Efimov states, the scaling limit
is practically approached as we see in Fig. 3. We expect that
going further in diminishing the absolute value of $E_2$ the new 
excited states will also follow the same route. The claim is of course
that the route is universal for all states in the scaling limit.

\begin{figure}[h]
\centerline{\epsfig{figure=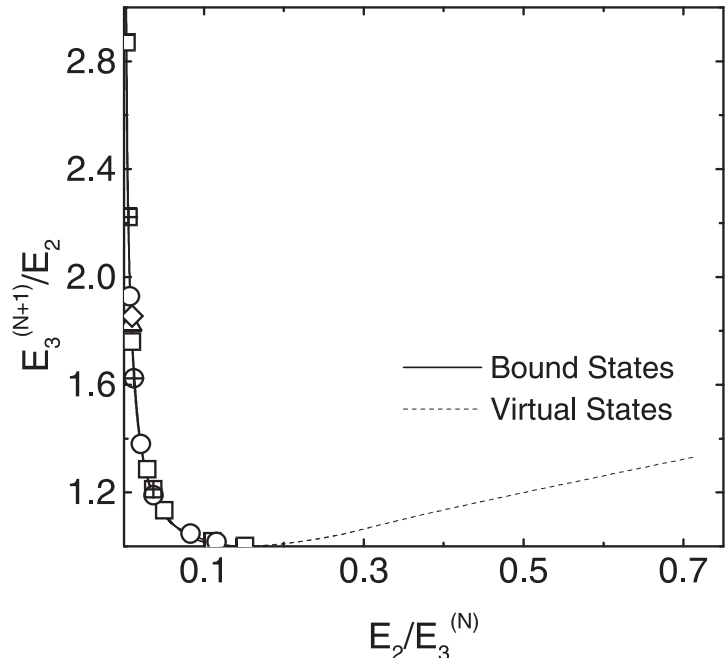,width=9.5cm}}
\caption[dummy]
{Ratio of the trimer excited or virtual $(N+1)-$th state
energy as a function of the ratio of  the dimer energy  and trimer $N$-th 
bound state energy. The results for the trimer excited bound state energies
are shown by the solid curve and the virtual state energies are
shown by the dotted curves. 
Our calculations show that the results for $N=0$ and $N=1$ practically 
coincide.
The symbols represent results from other calculations: 
empty squares ($s-$wave) and empty circles ($s+d$ waves)
are from Ref.~\cite{CG} (for $N=0$); crossed squares, from Ref.~\cite{huber3};
the crossed circle is from Ref.~\cite{BK}; the triangle, from Ref.~\cite{FJ};
and the lozenge, from Ref.~\cite{KMS}.}
\label{fig3}
\end{figure} 

The results for the energy of the excited Efimov state in
$^4$He$_3$ molecule given by ${\cal K}(z)$ 
($z=[E_2/E_3^{(N)}]^\frac12$), obtained by solving Eqs.(\ref{skt}),
(\ref{chi1}) and (\ref{vchi1}) in 
the scaling limit, are  compared to the realistic model calculations
also  presented in Fig. 4. The homogeneous integral equation
with the sharp cut-off momentum regulator, which generalizes
Eq.(\ref{skt}) for the virtual trimer state is not written explicitly
in the text as it can be easily derived.   
We observe the ratio
 $[({E^{(N+1)}_{3}}-E_2)/ {E^{(N)}_{3}}]^\frac12$ depends on 
$z$ for realistic models, as well. In this plot we only show
results for a bound dimer and $N=0$. 
The extreme limit of $z$ allowing the  excited  state are given by ${\cal 
K}(z)=0$ which gives $z=0.38$.
The solution of  Eqs.(\ref{skt}) and (\ref{chi1}) in the scaling limit 
qualitatively reproduces the results for  several interatomic potentials.
A deviation is seen for $z\approx 0.4$ that  are due to corrections from
the finite range of the potential.
The excited $(N+1)$ three-body state becomes virtual for 
$E_2/E_3^{(N)} > 0.145$ (as seen in Fig. 3), implying that 
$E_3^{(N)} < 6.9\;{\hbar^2}/{(m a^2)}$ in this case. 
This threshold value agrees with the value previously found in 
Refs.~\cite{am97,am99}, recently confirmed in Ref.~\cite{bra}, for 
the condition of the disappearance of the excited trimer state in the limit 
of a zero-range interaction.  
Let us stress that the regularization schemes used in Eqs.~(\ref{skt})
and (\ref{chi1}) are consistent not only for the calculation of the
bound excited trimer energies but also for the virtual trimer energies,
as  shown in Fig. 4. The small difference between the two regularization
schemes tends to vanish fast for higher values of $N$. 

\begin{figure}[h]
\centerline{\epsfig{figure=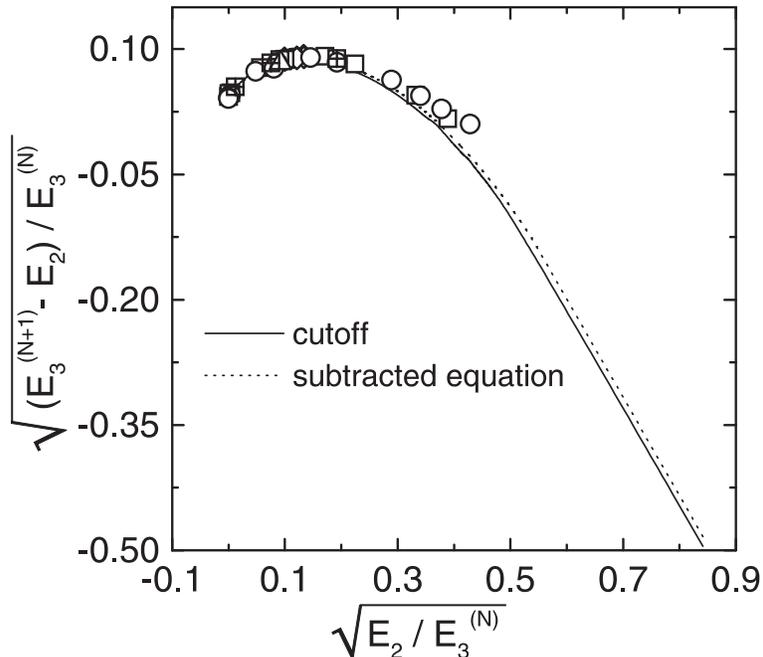,width=10.0cm}}
\caption{Results for the trimer bound and virtual excited
$(N+1)-$th state energies, scaled by the $N-th$ bound-state energy. 
A comparison between calculations performed with cutoff 
and subtraction methods for the regularizations is given for $N=0$. 
We also present results from other calculations, as described in the 
caption of Fig. 3.}
\label{fig4}
\end{figure} 

\section{Conclusions}

Natural scales determine the physics of quantum few-body systems with 
short-range interactions.  The physical scales of three interacting 
particles, in the state of zero total angular momentum, are identified 
with the bound or virtual subsystems energy and the ground state 
three-body binding energy. The scaling limit is found when 
the ratio  between the scattering length and the interaction range 
tends to infinity, while the ratio between the physical scales
are kept fixed. This  defines a scaling function for a given observable. 
From the formal point of view, we showed the relation of 
the scaling limit and  the renormalization aspects of a few-body model 
with a zero-range interaction, through the derivation of subtracted
three-body T-matrix equations, which are renormalization group invariant.

In the present work, we investigate the behavior of an
excited Thomas-Efimov state as the binding energy of the subsystem
increases with respect to the energy of the next lower bound three-body 
state. As shown, by allowing the two-body binding energy to increase 
in respect to the three-particle ground state energy, the excited 
three-body state disappears and a corresponding three-body virtual state 
emerges. The threshold for the three-body virtual state was found to be at the
energy  of the weakly bound trimer equal to $6.9\;{\hbar^2}/{(m a^2)}$ 
for large positive scattering lengths $(a)$.
The dependence of the $s-$wave virtual state three-body energy on 
the two and three atom ground state binding energies is calculated in the 
limit of a zero-range potential in a form of an universal scaling 
function. 
The scaling plots are an useful tool to classify observables 
and provide first guess to guide realistic calculations, as well as for 
planning experiments, with the aim of looking for weakly bound excited 
state of triatomic molecules.

The results of the present study can also be particularly relevant to
the interpretation of  experiments in atomic condensation, in which the 
effective atom-atom scattering length can be altered from negative to 
positive, in a wide range of values crossing zero-energy bound dimer~\cite{avaria}. 
For large positive scattering lengths, our estimate gives the threshold
for the zero-binding trimer state, which allows to settle the experimental 
conditions  for an investigation of the Efimov effect and search for
their influence on the observables of condensed systems. On the other hand,
large negative two-body scattering length have been 
recently investigated in Ref.~\cite{JILA02}. There is the possibility
that the observed discrepancy related with previous theoretical 
predictions can have their explanations in three-body effects, as well, 
because large two-body scattering lengths give the conditions where 
three-body (bound or virtual) Efimov states are likely to occur.  

We would like to thank Funda\c c\~ao de Amparo \`a 
Pesquisa do Estado de S\~ao Paulo (FAPESP) and Conselho Nacional de  
Desenvolvimento Cient\'{\i}fico e Tecnol\'ogico (CNPq) for
partial support.

\end{document}